\documentclass[aps,twocolumn,prd,showpacs,amsmath,amssymb]{revtex4}
\usepackage{dcolumn}
\usepackage{bm}
\usepackage{amsfonts}

\newcommand{\nn}{\nonumber\\}
\newcommand{\ep}{\epsilon}

\newcommand{\be}{\begin{equation}}
\newcommand{\bw}{\begin{widetext}}
\newcommand{\ew}{\end{widetext}}
\newcommand{\bse}{\begin{subequation}}
\newcommand{\ese}{\end{subequation}}
\newcommand{\ee}{\end{equation}} 
\newcommand{\eei}{\end{eqnarray}\indent\indent}
\newcommand{\bc}{\begin{center}}
\newcommand{\ec}{\end{center}}
\newcommand{\ber}{\begin{eqnarray}}
\newcommand{\eer}{\end{eqnarray}}
\newcommand{\ba}{\begin{array}}
\newcommand{\ea}{\end{array}}
\newcommand{\bal}{\begin{align}}
\newcommand{\eal}{\end{align}}

\newcommand{\sfrac}[2]{{\textstyle{#1\over#2}}}
\def\case#1/#2{\textstyle\frac{#1}{#2} }
\newcommand{\nb}{\nabla}
\newcommand{\D}{\tl\nb}

\newcommand{\tl}{\tilde}
\newcommand{\bver}{\begin{verbatim}}
\newcommand{\ever}{\end{verbatim}}

\begin{document}

\title{On shear-free perturbations of $f(R)$ gravity}
\author{Amare Abebe$^{1,2}$\footnote{amare.abebe@acgc.uct.ac.za}, Rituparno Goswami $^{1,2}$ \footnote{rituparno.goswami@uct.ac.za} and Peter K. S. Dunsby$^{1,2,3}$\footnote{peter.dunsby@uct.ac.za}}

\affiliation{1. Astrophysics, Cosmology and Gravity Centre (ACGC),
University of Cape Town, Rondebosch, 7701, South Africa}

\affiliation{2. Department of Mathematics and Applied Mathematics,
  University of Cape Town, 7701 Rondebosch, Cape Town, South Africa}

\affiliation{3. South African Astronomical Observatory,
  Observatory 7925, Cape Town, South Africa.}

\date{\today}

\begin{abstract}
Recently it was shown that if the matter congruence of a general relativistic perfect fluid flow in an almost FLRW universe is shear-free,  then it must be either expansion or rotation-free.  Here we generalize this result for a general $f(R)$ theory of gravity and show there exist scenarios where this result can be avoided. This suggests that there are situations where linearized forth-order gravity shares properties with Newtonian theory not valid in General Relativity. 
  \end{abstract}

\maketitle
\section{Introduction}

The observational evidence for the accelerated expansion of the universe, and the introduction of the concept of Dark Energy has put theoretical cosmology into crisis. This is due to the fact that despite an increasing amount and quality of data, no model has been proposed thus far that is able to give a completely satisfactory theoretical explanation of all the available observational data.  

Among the many different ways to achieve cosmic acceleration, the modification of the classical gravitational action based on General Relativity has recently gained much attention \cite{revnostra,Odintsov, Carroll,star2007,Sotiriou:2008rp,Clifton}. The reason for this popularity is due to the fact that these models provide a somewhat more natural explanation of the cosmic acceleration: this effect is due to corrections to Einstein gravity which are directly related to the characteristic properties of the gravitational interaction. Most investigations of higher order gravity have focused on Fourth Order Gravity (FOG), i.e., on gravitational Lagrangians in which the corrections are at most of order four in the metric,  and in what follows we will also focus on these models.

Because the field equations resulting from FOG are highly nonlinear,  difficult conceptual and technical issues arise which need to be resolved in order to uncover the detailed physics of these models.  Consequently it is crucial to develop new methods which are able to assist in resolving these problems. Two such approaches, the {\it dynamical systems approach to cosmology} and the {\it 1+3 covariant approach}, have proved very useful in contributing to our understanding of how the astrophysics and cosmology of these theories differ from what is found in General Relativity. Most of the work thus far has focussed on the dynamics of homogenous cosmological models \cite{homogeneous},  the linear growth of large scale structure \cite{perturbations} and on finding exact solutions which describe the gravitational field of stars and compact objects \cite{SS}. Of particular importance is to understand the relationship between the Newtonian and Relativistic limits of FOG which is important in describing the dynamics of nonlinear fluid flows in such theories. This is relevant both in the physics of gravitational collapse and the late (nonlinear) stages of structure formation \cite{newtonian}. 

Central to all of these problems is the differential properties of timelike geodesics which describe the fluid flow in cosmology. In general the  kinematics of such fluid flows are described by the expansion $\Theta$, shear (or distortion) $\sigma_{ab}$, rotation $\omega^c$, and acceleration $A_a$ of the four-velocity field $u^a$ tangent to the fluid flow lines, their governing equations obtained by contracting the Ricci identities (applied to $u^a$) along and orthogonal to $u^a$, which determine how they couple to gravity \cite{EllisCovariant}. 

The delicate relationship between the kinematic quantities in Newtonian and relativistic fluid flows in General Relativity is most strikingly seen in a remarkable result first obtained by Ellis in 1967 \cite{GFR}. In this paper it was found that {\em if the four velocity vector field of a barotropic perfect fluid with vanishing pressure is shear-free, then either the expansion or the rotation of the fluid vanishes}. This is a purely local result to which no corresponding Newtonian equivalent appears to hold,  as counter-examples can be explicitly constructed \cite{counter}. It is therefore interesting to ask whether such a result holds in the more general setting of FOG. 

As a first step towards this goal, we examine whether this result holds in situations where the hydrodynamic and gravitational equations have been linearized about a Friedmann-Lema\^{\i}tre-Robertson-Walker (FLRW) background \cite{EB,EBH,BDE,DBE,BED,DBBE}. These {\em almost FLRW} models can be thought of as lying somewhere between the full nonlinear situation and Newtonian theory, at least in the cosmological context, and therefore an analysis of the theorem in this context could shed some light on the generality of the result in FOG, indeed, since it has already been shown in earlier work that cosmologies with a bounce occur more naturally in such theories \cite{bounce}, one might expect a somewhat weaker version of the theorem to emerge. 

We show that if the 3-curvature vanishes, then  the result of \cite{GFR} can always be avoided for vacuum universes. We also demonstrate there is at least one physically realistic non-vacuum case in which both rotation and expansion are simultaneously possible.
\section{Field equations for $f(R)$ Gravity}
We know that for homogeneous and isotropic spacetimes, a ``sufficiently general" 
fourth-order Lagrangian only contains powers of  $R$ and we can write the action as 
\be
{\cal A}= \sfrac12 \int d^4x\sqrt{-g}\left[f(R)+2{\cal L}_m\right]\;,
\label{action}
\ee
where ${\cal L}_m$ represents the matter contribution.
Varying the action with respect to the metric gives the following field equations:
\be
f' G_{ab} = T^{m}_{ab}+ \sfrac12 (f-Rf')g_{ab} 
+ \nabla_{b}\nabla_{a}f'- g_{ab}\nabla_{c}\nabla^{c}f'\;,
\label{field1}	
\ee
where $f'$ denotes the derivative of the function $f$ w.r.t the 
Ricci scalar and  $T^{m}_{ab}$ is the matter stress energy tensor defined as
\be
T^{m}_{ab} = \mu^{m}u_{a}u_{b} + p^{m}h_{ab}+ q^{m}_{a}u_{b}+ q^{m}_{b}u_{a}+\pi^{m}_{ab}\;.
\ee 
Here $u^a$ is the direction of a timelike observer, 
$h^a_b=g^a_b+u^au_b$ is the projected metric on the 3-space perpendicular to $u^a$.  
Also $\mu^{m}$, $p^m$, $q^{m}$ and $\pi^m_{ab}$ denote the standard matter density, 
pressure, heat flux and anisotropic stress respectively. 
Equations (\ref{field1}) reduce to the standard Einstein field equations when $f(R) = R$.\\

We can write the modified field equations (\ref{field1}) as 
 \be
 G_{ab}=\tl T^{m}_{ab}+T^{R}_{ab}\equiv T_{ab}\;,
 \ee
where 
 \be
\tl T^{m}_{ab}=\frac{T^{m}_{ab}}{f'}\;,
\ee
\be
T^{R}_{ab}=\frac{1}{f'}\left[\sfrac{1}{2}(f-Rf')g_{ab}+\nb_{b}\nb_{a}f'-g_{ab}\nb_{c}
\nb^{c}f' \right]\,,
\ee
and the thermodynamic quantities of this fictitious ``{\it curvature fluid}'' 
are given by
\be
\mu^{R}=T^{R}_{ab}u^{a}u^{b}\;, ~p^R=\sfrac13h^{ab}T^{R}_{ab}\;,
\ee
\be
q_a^R=h_a^bT^{R}_{bc}u^c\;,~\pi^{R}_{ab}= \left[ h^{(a}{}_c {} h^{b)}{}_d 
- \sfrac{1}{3} h^{ab}h_{cd}\right] T^R_{cd}\;.
\ee
The {\it total} thermodynamic quantities are then written as
\ber
\mu\equiv\frac{\mu^{m}}{f'}+\mu^{R}\;,&&\;p\equiv\frac{p^{m}}{f'}+p^{R}\;,\nonumber\\
q_{a}\equiv \frac{q^{m}_{a}}{f'}+q^{R}_{a}\;,&&\;\pi_{ab}\equiv\frac{\pi^{m}_{ab}}{f'}+\pi^{R}_{ab}.
\eer

Using the standard  1+3 covariant approach \cite{EllisCovariant}, 
two derivatives are defined: the vector $ u^{a} $ is used to define the 
\textit{covariant time derivative} (denoted by a dot) for any tensor 
${T}^{a..b}{}_{c..d} $ along the observers' worldlines:
\be
\dot{T}^{a..b}{}_{c..d}{} = u^{e} \nb_{e} {T}^{a..b}{}_{c..d}~,
\ee
and the tensor $ h_{ab} $ is used to define the fully orthogonally 
\textit{projected covariant derivative} $\tl\nb$ for any tensor ${T}^{a..b}{}_{c..d} $:
\be
\tl\nb_{e}T^{a..b}{}_{c..d}{} = h^a{}_f h^p{}_c...h^b{}_g h^q{}_d 
h^r{}_e \nb_{r} {T}^{f..g}{}_{p..q}\;,
\ee
with total projection on all the free indices. 
Angle brackets denote orthogonal projections of vectors and the
orthogonally \textit{projected symmetric trace-free} PSTF part of tensors:
\be
V^{\langle a \rangle} = h^{a}{}_{b}V^{b}~, ~ T^{\langle ab \rangle} = \left[ h^{(a}{}_c {} h^{b)}{}_d 
- \sfrac{1}{3} h^{ab}h_{cd}\right] T^{cd}\;.
\label{PSTF}
\ee
This splitting of spacetime also naturally defines the 3-volume element
\be
\ep_{a b c}=-\sqrt{|g|}\delta^0_{\left[ a \right. }\delta^1_b\delta^2_c\delta^3_{\left. d \right] }u^d\;,
\label{eps1}
\ee
with the following identities:
\be
\ep_{a b c}\ep^{d e f}=3!h^d_{\left[ a \right. }h^e_bh^f_{\left. c \right] }\;,
\ep_{a b c}\ep^{d e c}=2!h^d_{\left[ a \right. }h^e_{\left. b \right] }\;.
\label{eps2}
\ee
The covariant derivative of the timelike vector $u^a$ can now be decomposed into the
irreducible parts as
\be
\nb_au_b=-A_au_b+\sfrac13h_{ab}\Theta+\sigma_{ab}+\ep_{a b c}\omega^c,
\ee
where $A_a=\dot{u}_a$ is the acceleration, $\Theta=\tl\nb_au^a$ is the expansion, 
$\sigma_{ab}=\tl\nb_{\langle a}u_{b \rangle}$ is the shear tensor and $\omega^{a}=\ep^{a b c}\tl\nb_bu_c$ 
is the vorticity vector. Similarly the Weyl curvature tensor can be decomposed 
irreducibly into the Gravito-Electric and Gravito-Magnetic parts as
\be
E_{ab}=C_{abcd}u^cu^d=E_{\langle ab\rangle}\;,~H_{ab}=\sfrac12\ep_{acd}C^{cd}_{be}u^e=H_{\langle ab\rangle}\;,
\ee
giving a covariant description of {\it Tidal forces} and {\it Gravitational radiation} respectively.
\section{Linearized Field equations about FLRW background}
In the 1+3 covariant perturbation theory \cite{EB,EBH,BDE,DBE,BED,Roy,DBBE,Roy2}, 
the quantities that vanish in the background spacetime are considered to be 
first order and are automatically  gauge-invariant by virtue of the Stewart and Walker 
lemma \cite{SW}. We consider  the background to be FLRW where the 
Hubble scale sets the characteristic scale of the perturbations. In the perturbed spacetime the 
standard matter is considered to be a perfect fluid  with the Energy Momentum tensor given by:
\be
T^m_{ab}=(\mu^m+p^m)u_au_b+p^mg_{ab}\;.
\label{EMT}
\ee
Furthermore, we assume  standard matter to have a barotropic linear equation of state 
$p^m=w\mu^m$  satisfying  the Weak and Dominant energy conditions:
\be
\mu^m>0\;\;;\;\;\mu^m+p^m>0\;\;;\;\;\mu^m\ge|p^m|\;.
\label{EC}
\ee
 Since the matter  is a perfect fluid the heat flux ($q^m_a$)  and the anisotropic stress 
($\pi^m_{a b}$) vanish in the perturbed spacetime.  In addition, since we consider 
shear-free perturbations the shear tensor $\sigma_{a b}$ vanishes identically.

For the {\it Curvature Fluid} the linearized thermodynamic quantities are given as
\be
\mu^{R}=\frac{1}{f'}\left[\sfrac{1}{2}(Rf'-f)-\Theta f'' \dot{R}+ f''\tilde{\nabla}^{2}R \right]\;,
\ee
\ber
p^{R}=\frac{1}{f'}\left[\sfrac{1}{2}(f-Rf')+f''\ddot{R}+f'''\dot{R}^{2}\right.\nonumber\\
\left.+\sfrac{2}{3}\left( \Theta f''\dot{R}-f''\tilde{\nabla}^{2}R \right) \right]\;,
\eer
\be
q^{R}_{a}=-\frac{1}{f'}\left[f'''\dot{R}\tilde{\nabla}_{a}R +f''\tilde{\nabla}_{a}\dot{R}-\sfrac{1}{3}f''\Theta \tilde{\nabla}_{a}R \right],
\label{qR} 
\ee
\be
\pi^{R}_{ab}=\frac{1}{f'}f''\tilde{\nabla}_{\langle a}\tilde{\nabla}_{b\rangle}R .
\label{piR} 
\ee
With the conditions above, the linearized field equations are then given by:
\subsection*{Propagation equations}
\be
\dot{\Theta}-\tl\nb_aA^a=-\sfrac13 \Theta^2-\sfrac12(\mu+3p)\;,
\label{R1}
\ee
\be 
\dot{\omega^{\langle a \rangle}}-\sfrac{1}{2}\ep^{abc}\tl\nb_bA_c=-\sfrac23\Theta\omega^a\;,
\label{R3}
\ee
\ber
\dot{E^{\langle a b\rangle}}-\ep^{cd\langle a}\tl\nb_cH^{\rangle b}_d=-\Theta E^{ab}-
\sfrac{1}{2}\dot{\pi}^{ab}_{R}\nonumber\\
-\sfrac{1}{2}\tl\nb^{\langle a}q^{b\rangle}_{R}-\sfrac{1}{6}\Theta\pi^{ab}_{R}\;,
\label{B1}
\eer
\be
\dot{H^{\langle ab \rangle}}+\ep^{cd\langle a}\tl\nb_cE^{\rangle b}_d=-\Theta H^{ab}+
\sfrac{1}{2}\ep^{cd\langle a}\tl\nb_c\pi^{\rangle b}_{d~R}\;,
\label{B2}
\ee
\be\label{B4}
\dot{\mu}_{m}=-(\mu_{m}+p_{m})\Theta,
\ee
\be
\dot{\mu}+\tl\nb^{a}q^{R}_{a}=-(\mu+p)\Theta;
\ee
\subsection*{Constraint equations}
\be
(C_0)^{ a b}:=E^{a b}-\tl\nb^{\langle a}A^{b \rangle}-\sfrac{1}{2}\pi_{R}^{ab}=0\;,
\label{R2}
\ee
\be
(C_1)^a:=\tl\nb^a\Theta-\sfrac{3}{2}\ep^{abc}\tl\nb_b\omega_c-\sfrac{3}{2}q^{a}_{R}=0\;,
\label{R4}
\ee
\be 
(C_2):=\tl\nb^a\omega_a=0\;,
\label{R5}
\ee
\be
(C_3)^{ a b}:=H^{a b}+\tl\nb^{\langle a}\omega^{b \rangle}=0\;.
\label{R6}
\ee
\be
(C_4)^a:=\tl\nb^ap_{m} +(\mu_{m}+p_{m}) A^a=0\;,
\label{B3}
\ee
\be
(C_5)^a:=\tl\nb_bE^{a b}+\sfrac{1}{2}\tl\nb_{b}\pi^{ab}_{R}-\sfrac13\tl\nb^a\mu+
\sfrac13\Theta q^{a}_{R}=0\;,
\label{B5}
\ee
\be
(C_6)^a:=\tl\nb_bH^{a b}+(\mu+p)\omega^a+\sfrac12\ep^{abc}\tl\nb_{b}q_{c}^{R}=0\;.
\label{B6}
\ee
We note that the constraints $(C_1)^a$, $(C_2)$, $(C_3)^{a b}$, $(C_5)^a$ and $(C_6)^a$ 
are the constraints of Einstein field equations for general matter motion and are shown 
to be consistently {\it time propagated} along $u^a$ locally in General Relativity. 
However, the conditions 
$\sigma_{a b}=0$  and $q^a_{m}=0$ give the two new constraints $(C_0)^{a b}$ and $(C_4)^a$ 
respectively. 
In what follows we will use the following linearized commutation relations for shear-free congruences. 
For any scalar $\phi$
\ber
[\tl\nb_a\tl\nb_b-\tl\nb_b\tl\nb_a]\phi&=&2\ep_{a b c}\omega^c\dot \phi \;, \nonumber\\ 
\ep^{a b c}\tl\nb_b\tl\nb_c \phi&=&2\omega^a \dot \phi\;.
\label{C1}
\eer
If the gradient of the scalar is of the first order, we have 
\ber
[\tl\nb^a\tl\nb_b\tl\nb_a-\tl\nb_b\tl\nb^2]\phi&=&\sfrac{1}{3}\tl{R}\tl\nb_{b}\phi
\label{C2}
\eer
and
\ber
[\tl\nb^2\tl\nb_b-\tl\nb_b\tl\nb^2]\phi&=&\sfrac{1}{3}\tl{R}\tl\nb_{b}\phi+2\ep_{dbc}\tl\nb^d(\omega^c\dot \phi),
\label{C3}
\eer
where $\tl{R}=2\left(\mu-\frac13\Theta^2\right)$ is the 3-curvature scalar.
Also for any first order 3-vector $V^a=V^{\langle a \rangle}$, we have
\ber
[\tl\nb^a\tl\nb_b-\tl\nb_b\tl\nb^a]V_a&=&\sfrac13\tl{R}h^a{}_{\left[a\right. }
V_{\left. b \right]}\;,
\label{C4}
\eer
\ber
h^{a}{}_{c}h^{d}{}_{b}(\tl\nb_dV^c)\dot{}=\tl\nb_b\dot{V^{\langle a \rangle}}-\sfrac{1}{3}\Theta \tl\nb_bV^a,
\label{C6}
\eer
\ber
h^{a}{}_{c}(\tl\nb^2V^c)\dot{}=\tl\nb_b(\tl\nb^{\langle b}V^{a \rangle})\dot{}-\sfrac{1}{3}\Theta \tl\nb^2V^a.
\label{C5}
\eer
\section{Consistency of the new constraints}
We have already seen that the conditions of shear-free perturbations together with the matter 
being described by a perfect fluid in the perturbed spacetime, gives the new constraints 
$(C_0)^{a b}$ and $(C_4)^a$ respectively. To check their compatibility with the 
existing constraints of Einstein's field equations, we substitute $(C_0)_{b d}$ 
into $(C_5)_b$ to obtain
\be 
\tl\nb^d\tl\nb_{\langle b}A_{d \rangle}-\sfrac{1}{3}\tl\nb_b\mu+\tl\nb^{d}\pi^{R}_{bd}+\sfrac{1}{3}\Theta q^{R}_{b}=0\;.
\label{subs1}
\ee
Now from the constraint $(C_4)_b$ we have 
\be
A_b=-\frac{w}{w+1}\tl\nb_b\phi\;,
\label{subs2}
\ee
where $\phi=\ln(\mu_{m})$. Using equation (\ref{subs2}) in (\ref{subs1}) we get the 
constraint
\be
\frac{w}{w+1}\tl\nb^d\tl\nb_{\langle b}\tl\nb_{d \rangle}\phi+\sfrac{1}{3}\tl\nb_b\mu-\tl\nb^{d}\pi^{R}_{bd}-\sfrac{1}{3}\Theta q^{R}_{b}=0.
\label{newcons}
\ee
We note that for the new constraints to be compatible with the existing ones 
the above constraint must be satisfied. To check the spatial consistency of the 
above constraint on any initial hypersurface we take the curl of (\ref{newcons}) 
to get
\ber
&&\frac{w}{w+1}\ep^{acb}\tl\nb_c \tl\nb^d\tl\nb_{\langle b}\tl\nb_{d \rangle}\phi+\sfrac{1}{3}\ep^{acb}\tl\nb_c\tl\nb_b\mu\nn
&&-\ep^{acb}\tl\nb_c\tl\nb^{d}\pi^{R}_{bd}-\sfrac{1}{3}\Theta \ep^{acb}\tl\nb_c q^{R}_{b}=0\;,
\label{curl1}
\eer
which, on using (\ref{C1}) gives
\ber
&&\frac{w}{w+1}\ep^{acb}\tl\nb_c \tl\nb^d\tl\nb_{\langle b}\tl\nb_{d \rangle}\phi+\sfrac{2}{3}\omega^a\dot{\mu}\nn
&&+\sfrac{1}{3}\Theta \ep^{acb}\tl\nb_c \left[\frac{f'''}{f'}\dot{R}\tl\nb_{b}R+\frac{f''}{f'}\tl\nb_{b}\dot{R}-\frac{\Theta f''}{3f'}\tl\nb_{b}R\right]\nn
&&-\ep^{acb}\tl\nb_c\tl\nb^{d}\left[\frac{f''}{f'}\tl\nb_{\langle b}\tl\nb_{d\rangle}R\right]=0.
\eer
Breaking the PSTF part according to equation (\ref{PSTF}), using the commutators
(\ref{C2}), (\ref{C3}) and keeping only terms up to first order, we have:
\ber
&&\frac{w}{w+1}\ep^{acb}\left[\sfrac{2}{3}\tl\nb_c\tl\nb_b\tl\nb^2\phi+\sfrac{1}{3}\tl{R}\tl\nb_c\tl\nb_b\phi+\dot{\phi}\ep_{dbk}\tl\nb_c\tl\nb^d\omega^k\right]\nn
&&+\sfrac{2}{3}\omega^a\dot{\mu}-\frac{f''}{f'}\ep^{acb}\tl\nb_c\tl\nb^{d}\tl\nb_{\langle b}\tl\nb_{d\rangle}R+\frac{f'''\Theta\dot{R}}{3f'}\ep^{acb}\tl\nb_{c}\tl\nb_{b}R\nn
&&+\frac{f''\Theta}{3f'}\ep^{acb}\tl\nb_{c}\tl\nb_{b}\dot{R}-\frac{f''\Theta^{2}}{9f'}\ep^{acb}\tl\nb_{c}\tl\nb_{b}R=0.
\eer
Again using (\ref{C1}) and (\ref{eps2}) in the above equation and linearizing 
we get
\ber
&&\frac{w}{w+1}\left[\sfrac{2}{3}\tl{R}\omega^a\dot{\phi}-\dot{\phi}\tl\nb_k\tl\nb^a\omega^k+ \dot{\phi}\tl\nb^2\omega^a\right]+\sfrac{2}{3}\omega^a\dot{\mu}\nn
&&-\frac{f''}{f'}\ep^{acb}\left[\sfrac{2}{3}\D_{c}\D_{b}\D^{2}R+\sfrac{1}{3}\tl{R}\D_{c}\D_{b}R+\dot{R}\ep_{dbk}\D_{c}\D^{d}w^{k}\right]\nn
&&+\frac{2\omega^{a}}{3f'}\left[f'''\Theta\dot{R}^{2}-\sfrac{1}{3}f''\Theta^{2}\dot{R}+f''\Theta\ddot{R}\right]=0.
\eer
This can also be written as
\ber
&&\frac{w}{w+1}\left[\sfrac{2}{3}\tl{R}\omega^a\dot{\phi}-\dot{\phi}\tl\nb_k\tl\nb^a\omega^k+ \dot{\phi}\tl\nb^2\omega^a\right]+\sfrac{2}{3}\omega^a\dot{\mu}\nn
&&- \dot{R}\frac{f''}{f'}\left[\sfrac{2}{3}\tl{R}\omega^{a}-\D_{k}\D^{a}\omega^{k}+\D^{2}\omega^{a}\right]\nn
&&+\frac{2\omega^{a}}{3f'}\left[f'''\Theta\dot{R}^{2}-\sfrac{1}{3}f''\Theta^{2}\dot{R}+f''\Theta\ddot{R}\right]=0.
\eer
Now, from relation (\ref{C3}) and using (\ref{R5}) we know that
\be
\tl\nb_k\tl\nb^a\omega^k=\sfrac{1}{3}\tl{R}\omega^a\;,
\ee
and from (\ref{B4}) we have 
\be
\dot{\phi}=-(1+w)\Theta\;.
\label{curl5}
\ee
Thus rearranging terms gives
\ber
&&w\Theta\left[\frac{\tl{R}}{3}\omega^a+ \tl\nb^2\omega^a\right]+\sfrac{2}{3}(\mu+p)\Theta\omega^a
+ \dot{R}\frac{f''}{f'}\left[\frac{\tl{R}}{3}\omega^{a}\right.\nn
&&\left.+\D^{2}\omega^{a}\right]-\frac{2\Theta \omega^{a}}{3f'}\left[f'''\dot{R}^{2}-\sfrac{1}{3}f''\Theta\dot{R}+f''\ddot{R}\right]=0\;.\nonumber\\
\eer
Since \ber
\mu+p=\frac{(1+w)\mu_{m}}{f'}-\frac{f''\dot{R}\Theta}{3f'}+\frac{f''\ddot{R}}{f'}\nn
+\frac{f'''\dot{R}^{2}}{f'}+\frac{f''}{3f'}\D^{2}R,
\eer
the above equation, to linear order,  simplifies to
 \ber
&&w\Theta\left[\frac{\tl{R}}{3}\omega^a+ \tl\nb^2\omega^a\right]+\frac{2(1+w)\mu_{m}\Theta\omega^a}{3f'}\nn
&&+ \dot{R}\frac{f''}{f'}\left[\sfrac13{\tl{R}}\omega^{a}+\D^{2}\omega^{a}\right]=0.
\eer
Further manipulation leads to
\ber\label{conricc}
&&\omega^{a}\left[ \left(\frac{w\Theta}{3}+\frac{\dot{R}f''}{3f'}\right)\tl{R}
+\frac{2(1+w)\mu_{m}\Theta}{3f'}\right]\nn
&&+\left(\frac{\dot{R}f''}{f'}+w\Theta\right)\D^{2}\omega^{a}=0.
\eer
We know that in terms of the scale factor $a(t)$ of a FLRW spacetime, the 
expansion, acceleration, jerk and snap parameters are defined by the following relations:
\ber
\Theta=3\frac{\dot a}{a}\;, ~~~~&&\; q=-\frac{\ddot{a}a}{\dot{a}^{2}},\\
j=\frac{\dddot{a}a^{2}}{\dot{a}^{3}}\;,~~~~&&\;s=\frac{a^{3}}{\dot{a}^{4}}\frac{d^{4}a}{dt^{4}}.
\eer
From the above equations we can easily see that the time propagations of these 
quantities can be written as
\ber
&&\dot{\Theta}=-\frac{1}{3}\Theta^{2}(1+q)\;,\\
&&\dot{q}=-\frac{1}{3}\Theta\left(j-q-2q^{2}\right)\;,\\
&&\ddot{\Theta}=\frac{1}{9}\Theta^{3}\left(2+3q+j\right)\;,\\
&&\dot{j}=\frac{1}{3}\Theta\left(s+2j+3qj\right)\;,\\
&&\ddot{q}=-\frac{1}{9}\Theta^{2}\left[s+2j-3q^{2}+6qj-6q^{3}\right]\;.
\eer
Then, the Ricci scalar $R$  is given by
\be
R=\frac{2}{3}\Theta^{2}(1-q)+\tl{R}
\ee
and hence
\be
\dot{R}=\frac{2}{3}\Theta Q\;,
\ee
where
\ber
&&Q=\frac{1}{3}\Theta^{2}(j-q-2)+\tl{R},\\
&&\dot{Q}=\sfrac{1}{9}\Theta\left[(4+5q+j+jq+s)\Theta^{2}+6\tl{R}\right]\;.
\label{dop}
\eer
This means that we can rewrite (\ref{conricc}) as
\ber\label{conric}
&&\sfrac{2}{3}\Theta\bigg\{\omega^{a}\left[ \left(\frac{w}{2}+\frac{f''}{3f'}Q\right)\tl{R}+\frac{(1+w)\mu_{m}}{f'}\right]\nn
&&+\left[\frac{f''}{f'}Q+\frac{3w}{2}\right]\D^{2}\omega^{a}\bigg\}=0\;.
\eer
We can see from this equation that spatial consistency requires the vanishing of  either $\Theta$ or the terms in the curly brackets. 

To check for temporal consistency of the new constraint (\ref{conric})  we take its time evolution which can be written as
\ber
&&\omega^{a}\bigg\{\Theta\left[\frac{(1-w)P}{3}\tl{R}+\frac{(1+w)}{f'}\frac{(3w+5)f'+4f''Q}{6f'}\mu_{m}\right]\nn
&&+\frac{\dot{P}}{P}\left[(\frac{1+w}{f'})\mu_{m}\right]\bigg\}=0\;,
\eer
where we have used 
\ber
\dot{\omega}^{a}=(w-\sfrac{2}{3})\Theta\omega^{a}
\eer
and
\ber
(\D^{2}\omega^a)^{.}=\frac{(3w-5)}{6}\Theta\D^{2}\omega^a+\frac{(w-1)}{6}\Theta\tl{R}\omega^a\;.\eer
We have also defined
\be P\equiv  \frac{f''}{f'}Q+\frac{3w}{2}\;.\ee

From (\ref{dop}), we can write
\be
\dot{P}=Z\Theta\;,
\ee
where
\ber
&&Z=\frac{2}{3}\left(\frac{f'''}{f'}-(\frac{f''}{f'})^{2}\right)Q^{2}\nn
&&+\frac{f''}{9f'}\left((4+5q+j+jq+s)\Theta^{2}+6\tl{R}\right).
\eer
Equation  (\ref{conric}) can then be rewritten as
\ber\label{conricci}
&&\Theta\omega^{a}\bigg\{\left[\frac{(1-w)P}{3}\tl{R}+\frac{(1+w)}{f'}\frac{(3w+5)f'+4f''Q}{6f'}\mu_{m}\right]\nn
&&+\frac{Z}{P}\left[(\frac{1+w}{f'})\mu_{m}\right]\bigg\}=0\;.
\eer
It follows that for the new constraints to be spatially  and temporally consistent we must have either $\omega^a\Theta=0$ or
the expression in the curly brackets must vanish.  It is interesting to see whether there exist solutions  of a given $f(R)$ theory of gravity which can avoid the Ellis condition. 

From (\ref{conricci}), it is easy to see that if the 3-curvature vanishes,  then  the result of ~\cite{Nzioki} can always be avoided for vacuum universes ($\mu_m=0$). This implies that {\em a shear-free, spatially flat vacuum universe in any $f(R)$ theory can rotate and expand  simultaneously in the linearized regime.} 

The non-vacuum case is more difficult to analyze in general;  however, as we will see below there does exist at least one non-trivial case which does violate the Ellis condition.

For a flat Milne universe, where the  matter energy density is given by 
$\mu_{m}=\frac{\mu_{0}}{ a^{3(1+w)}}$, we have
\ber
&&\dot{\Theta}=-\sfrac{1}{3}\Theta^{2},\\
&&R=\sfrac{2}{3}\Theta^{2},\\
&&a(R)=\frac{1}{\sqrt{R}},\\
&&\dot{R}=-\sqrt{\sfrac{2}{3}}R^{\frac{3}{2}}\,.
\eer
Substituting these quantities into the Friedmann equation
\be
\frac{1}{3}\Theta^{2}=\frac{1}{f'}\left[\mu_{m}+\frac{Rf'-f}{2}-\Theta\dot{R}f''\right]\;,
\ee
one gets
\be
-R^2\frac{d^{2}f(R)}{dR^{2}}+\frac{f(R)}{2}-\frac{\mu_{0}}{a(R)^{3(1+w)}}=0\;,
\ee
which has the following general solution:
\be\label{sol1}
f(R)=C_{1}R^{\frac{1+\sqrt{3}}{2}}+C_{2}R^{\frac{1-\sqrt{3}}{2}}-
\frac{4\mu_{0}}{1+12w+9w^{2}}R^{\frac{3(1+w)}{2}}.
\ee
Let us only consider the particular solution (the last term of the above equation), 
which is an $R^n$ - theory of gravity.
Now,  if we look at (\ref{conricci}), for the corresponding flat Milne universe in 
$R^{n}$ gravity, the term in the curly brackets reduces to
\be\label{sol2}
\frac{(1+w)\mu_{m}}{6f'}\left[3w+9-4n\right]=0.
\ee
Comparing solutions (\ref{sol2}) and the particular solution of 
(\ref{sol1}) (with $n=3(1+w)/2$) we get $w=1$ if $\mu_{m}\ne0$. In other words, 
{\em for a stiff fluid in $R^3$ gravity, there exists a flat Milne-universe solution 
which can rotate and expand simultaneously at the level of linearized perturbation theory.}
\section{Discussion and Conclusion}
In this paper we consider shear-free fluid flows in $f(R)$ gravity in situations where the hydrodynamic and gravitational equations have been linearized about a FLRW  background. This extends recent work by Nzioki {\it et. al.}, which considered such situations in General Relativity \cite{Nzioki}.   We showed that if the 3-curvature vanishes, then  the result of \cite{GFR, JMM} can always be avoided for vacuum universes. We also demonstrated there is at least one physically realistic non-vacuum case in which both rotation and expansion is simultaneously possible. This suggests that there are situations where linearized forth-order gravity shares properties with Newtonian theory not valid in General Relativity.


\begin{thebibliography}{99}
\bibitem{revnostra}
 S.~Capozziello, S.~Carloni and A.~Troisi,
   ``Recent Research Developments in Astronomy \&  Astrophysics"-RSP/AA/21 (2003); S.~Capozziello, V.~F.~Cardone, S.~Carloni and A.~Troisi,  Int.\ J.\ Mod.\ Phys.\ D {\bf 12} (2003) 1969; S.~Capozziello, V.~F.~Cardone and A.~Troisi,  JCAP {\bf 0608} (2006) 001; S.~Capozziello,  Int.\ J.\ Mod.\ Phys.\  D {\bf 11}, 483 (2002).
 \bibitem{Odintsov}
  S.~Nojiri and S.~D.~Odintsov,  Phys.\ Rev.\  D {\bf 68}, 123512  (2003).
 \bibitem{Carroll}
 S.~M.~Carroll, V.~Duvvuri, M.~Trodden and M.~S.~Turner,
  Phys.\ Rev.\  D {\bf 70}, 043528 (2004).
\bibitem{star2007} A.~A.~Starobinsky,  JETP Lett.\  {\bf 86}, 157 (2007). 
\bibitem{Sotiriou:2008rp}
 T.~P.~Sotiriou and V.~Faraoni, arXiv:0805.1726 [gr-qc].
 \bibitem{Clifton}
  T.~Clifton, P.~G.~Ferreira, A.~Padilla and C.~Skordis,
  arXiv:1106.2476 [astro-ph.CO].
\bibitem{homogeneous}
 L.~Amendola, R.~Gannouji, D.~Polarski and S.~Tsujikawa,
  Phys.\ Rev.\  D {\bf 75}, 083504 (2007), S.~Carloni, P.~K.~S.~Dunsby, S.~Capozziello and A.~Troisi, Class.\ Quant.\ Grav.\  {\bf 22}, 4839 (2005);  S.~Carloni, A.~Troisi and P.~K.~S.~Dunsby, Gen.\ Rel.\ Grav.\  {\bf 41}, 1757 (2009);  J.~A.~Leach, S.~Carloni and P.~K.~S.~Dunsby,   Class.\ Quant.\ Grav.\  {\bf 23}, 4915 (2006);  N.~Goheer, J.~Larena and P.~K.~S.~Dunsby,  Phys.\ Rev.\  D {\bf 80}, 061301 (2009);  P.~K.~S.~Dunsby, E.~Elizalde, R.~Goswami, S.~Odintsov and D.~S.~Gomez,  Phys.\ Rev.\  D {\bf 82}, 023519 (2010).
\bibitem{perturbations}
  R.~Bean, D.~Bernat, L.~Pogosian, A.~Silvestri and M.~Trodden,
 Phys.\ Rev.\  D {\bf 75}, 064020 (2007);    Y.~S.~Song, W.~Hu and I.~Sawicki,
  Phys.\ Rev.\  D {\bf 75}, 044004 (2007); S.~Carloni, P.~K.~S.~Dunsby and A.~Troisi,  Phys.\ Rev.\  D {\bf 77}, 024024 (2008);  K.~N.~Ananda, S.~Carloni and P.~K.~S.~Dunsby,  Class.\ Quant.\ Grav.\  {\bf 26}, 235018 (2009).
\bibitem{SS}
 T.~Clifton, Class.\ Quant.\ Grav.\  {\bf 23}, 7445 (2006); A.~M.~Nzioki, S.~Carloni, R.~Goswami and P.~K.~S.~Dunsby,  Phys.\ Rev.\  D {\bf 81}, 084028 (2010).
 \bibitem{newtonian} 
S. Matarrese, O. Pantano and D. Saez, Phys. Rev. Lett. {\bf 72},  320 (1994).
\bibitem{EllisCovariant} G.~F.~R.~Ellis \& H van Elst,
``Cosmological Models", Carg\`{e}se Lectures 1998, in Theoretical
and Observational Cosmology, Ed. M Lachi�ze-Rey, (Dordrecht: Kluwer,
1999), 1. [arXiv:gr-qc/9812046].
\bibitem{GFR} G. F.R. Ellis, J. Math. Phys. {\bf 8}, 1171-1194, (1967).
\bibitem{counter} O. Heckmann and E. Sch\"{u}cking,  Handbuch der Physik LIII Edited by
S. Fl\"{u}gge. (Springer Verlag, Berlin-G\"{o}tingin-Heidelberg) 489 (1959).
\bibitem{EB} G.~F.~R.~Ellis \& M.~Bruni, Phys Rev D {\bf 40} 1804 (1989).
\bibitem{EBH} G.~F.~R.~Ellis, M.~Bruni and J.~Hwang,
Phys.\ Rev.\  D {\bf 42} 1035 (1990).
\bibitem{BDE}
M.~Bruni,  P.~K.~S.~Dunsby \& G.~F.~R.~Ellis,
Ap. J. {\bf 395} 34 (1992).
\bibitem{DBE}  P.~K.~S.~Dunsby, M.~Bruni and G.~F.~R.~Ellis,
Astrophys.\ J.\  {\bf 395}, 54 (1992).
\bibitem{BED} M.~Bruni, G.~F.~R.~Ellis and P.~K.~S.~Dunsby,
Class.\ Quant.\ Grav.\  {\bf 9}, 921 (1992).
\bibitem{DBBE}   P.~K.~S.~Dunsby, B.~A.~C.~Bassett and G.~F.~R.~Ellis,
Class.\ Quant.\ Grav.\  {\bf 14}, 1215 (1997).
 \bibitem{bounce}
 S.~Carloni, P.~K.~S.~Dunsby and D.~M.~Solomons,  Class.\ Quant.\ Grav.\  {\bf 23}, 1913 (2006).
\bibitem{Roy}
R.~Maartens and J. ~Triginer, 
Phys.\ Rev.\  D {\bf 56} 4640 (1997).
\bibitem{Roy2}
R.~Maartens, T. ~Gebbie and  G.~F.~R.~Ellis, 
Phys.\ Rev.\  D {\bf 59} 083506 (1999).
\bibitem{SW}J.  M. Stewart and M.  Walker, Proc. R. Soc. London A341, 49 (1974).
 \bibitem{Nzioki}
  A.~M.~Nzioki, R.~Goswami, P.~K.~S.~Dunsby and G.~F.~R.~Ellis,  arXiv:1107.5410 [gr-qc] (2011).
  \bibitem{JMM} J. M. M Senovilla, C. F. Sopuerta and P. Szekeres, Gen. Rel. Grav. {\bf 30}, 389-411, (1998);  Phys Rev D {\bf 40} 1804 (1989).
  \end{thebibliography}
\end{document}